\documentclass[onecolumn,authoryear]{els-mrw} 

\usepackage{amsmath,amssymb,amsfonts,amsthm,makeidx,graphicx}
\usepackage{txfonts}
\usepackage{helvet}
\usepackage{soul}

\usepackage{amsmath,amssymb}
\usepackage{graphicx}
\usepackage[table]{xcolor} 
\usepackage{multirow} 
\usepackage{colortbl}
\usepackage{natbib}

\usepackage{aas_macros}
\begin{document}
\chapter{Evolution of Triple Stars}\label{chap1}

\author[1]{Hagai B. Perets}
\address[1]{\orgname{Technion - Israel Institute of Technology}, \orgdiv{Department of Physics}, \orgaddress{Technion city, Haifa, Israel 3200002}}

\articletag{Chapter Article tagline: update of previous edition,, reprint..}

\maketitle



\begin{abstract}[Abstract]
Triple stellar systems, consisting of three gravitationally bound stars, play a fundamental role in a wide array of astrophysical processes, from stellar evolution to the formation of exotic objects and gravitational wave sources. This review provides a comprehensive overview of the dynamics and evolution of triple stellar systems, highlighting their crucial role in shaping stellar populations and driving diverse astrophysical phenomena. We begin by discussing the observed properties of triples, including their frequency and orbital configurations, emphasizing the challenges in characterizing these systems. We then delve into the intricate dynamics of triples, exploring hierarchical, secular, quasi-secular, and chaotic regimes, with particular attention to the von Zeipel-Lidov-Kozai (vZLK) mechanism and its extensions. The interplay between stellar evolution and triple dynamics is examined, including mass loss, mass transfer, common-envelope evolution, and supernovae effects. We discuss the development and applications of triple population synthesis codes, demonstrating their power in modeling complex evolutionary scenarios and predicting observational outcomes. Finally, we emphasize the unique role of triple evolution in producing a wide range of astrophysical phenomena, from interacting binaries and stellar mergers to explosive transients and gravitational wave sources, underscoring the importance of triple systems in our understanding of stars and stellar populations.
\end{abstract}

\begin{BoxTypeA}[chap1:box2]{\large{Key Points}}

\begin{itemize}
\item Triple stellar systems are abundant and play a crucial role in various astrophysical phenomena.
\item The dynamics of triple systems are complex and can be categorized into hierarchical, secular, quasi-secular, and chaotic regimes.
\item The von Zeipel-Lidov-Kozai (vZLK) mechanism and its extensions are key drivers of eccentricity and inclination variations in hierarchical triples.
\item Stellar evolution processes, such as mass loss and mass transfer, significantly influence the dynamical evolution of triple systems.
\item The complex interplay between stellar evolutionary processes, dissipative processes from stellar interactions and the gravitational dynamics of triple systems determine their evolution and their byproducts. 
\item Triple evolution can lead to a wide range of outcomes, including close binary interactions, stellar mergers, the formation of compact object binaries, and the production of high energy transients and gravitational waves sources.
\end{itemize}
\end{BoxTypeA}

\section{Introduction}

Triple stellar systems, comprising three gravitationally bound stars, constitute a significant and ubiquitous component of the stellar population. Observational surveys reveal that a substantial fraction of stellar systems, particularly those containing massive stars, originate in triple or higher-order configurations (see Table 1). These systems, encompassing a broad spectrum of stellar masses from low-mass M dwarfs to massive O-type stars, play a pivotal role in various astrophysical phenomena. Their influence extends to the formation of compact and interacting binaries, and they contribute significantly to the genesis of black holes, neutron stars, and progenitors of stellar explosions and mergers \citep[e.g][]{Too+20}. Consequently, a thorough understanding of the dynamics and evolution of triple systems, and their interplay with dissipative and stellar evolutionary processes, is indispensable for a comprehensive understanding of stellar evolution and its implications for astrophysical phenomena.

The dynamics of triple systems exhibit considerably greater complexity compared to binary systems, primarily due to the hierarchical nature of their orbital configurations. Typically, a triple system consists of an inner binary orbited by a more distant tertiary companion (where the the teriary, together with the inner binary center-of-mass are termed the outer-binary in the triple. This hierarchical structure gives rise to a diverse range of dynamical behaviors, spanning from long-term secular evolution\footnote{The evolution is considered to be secular if it occurs on timescales much longer than the orbital period of the outer binary of a triple.} governed by mechanisms such as von Zeipel-Lidov-Kozai (vZLK) oscillations \citep[see][for a review]{Naoz2016}, to quasi-secular\footnote{The evolution is considered to be qausi-secular if it occurs on timescales much longer than the orbital period of the inner binary, but could be comparable to the orbit of the outer period.} and chaotic interactions that can culminate in stellar collisions or ejections. These multifaceted processes can drive the triple components into intense dynamical and dissipative interactions, encompassing mass loss, mass transfer, and various dissipative processes, thereby profoundly influencing the system's long-term evolution through close encounters, ejections, and even mergers.

\section{Definition and Structure of Triple Stellar Systems}

A triple-stellar system comprises three stars bound by mutual gravitational attraction, and engaged in long-term interactions. These systems typically exhibit a hierarchical configuration, wherein two stars form a close binary pair (the inner binary, whose properties are typically denoted by \emph{in} or the index 1 throughout), while the third star orbits this binary at a significantly greater distance (the outer companion; outer binary, typically denoted by \emph{out} or the index 2 throughout). This hierarchical arrangement ensures the system's long-term stability, as the gravitational influence of the outer star is sufficiently weak to avoid disrupting the inner binary on short timescales.

Although all orbital parameters of a hierarchical triple system contribute to its dynamics, the key parameters governing its type of evolution include the masses of the three constituent stars ($m_1$, $m_2$, $m_3$).  the semi-major axes ($a_{\mathrm{in}}$, and $a_{\mathrm{out}}$) of the inner and outer orbits, and the eccentricities ($e_{\mathrm{in}}$ and $e_{\mathrm{out}}$) of these orbits.  The mutual inclination between the inner and outer orbits ($i$) plays a crucial role in the system's long-term evolution, particularly in secular processes such as the von Zeipel-Lidov-Kozai mechanism, which can induce substantial changes in the eccentricity and inclination of the inner binary, as elaborated upon subsequently. While stable, long-lived triple systems conserve their semi-major axes, the eccentricities and mutual inclination can undergo alterations due to secular and quasi-secular processes. 

We also introduce additional orbital parameters, used below; these include the argument of the periapsis periastron $\omega$; $\omega_1$ and $\omega_2$ for the inner and outer orbits ) which is the angle between the ascending node and the periapsis (the point of closest approach between the two bodies in the orbit), measured in the orbital plane. The longitude of the ascending node ($\Omega$; $\Omega_1$ and $\Omega_2$ for the inner and outer orbits ) is the angle between a reference direction and the ascending node (where the orbit crosses the reference plane), measured in the reference plane.  We should also mention the total angular momentum, $j$ and the angular momentum in the z-direction, $j_z$; the former is a conserved quantity, and the latter is conserved only in some parameter regimes. 

\section{Observed Properties of Triple Systems}
The statistical properties of triple stellar systems exhibit a clear dependence on the stellar mass and the stellar type. While extensively investigated, the inherent challenges in identifying all stellar companions and/or accurately measuring all orbital parameters of such systems impose limitations on our current knowledge of their detailed properties. This knowledge remains far more constrained than that of single or binary systems.

Specifically, a diverse range of detection and characterization techniques is necessary to encompass the full spectrum of stars and orbital ranges. These techniques span from spectroscopic analysis and photometry (through potential eclipses) for short-period orbits, to interferometric surveys for wider orbits, and astrometric measurements for the widest separations \citep{Moe+17,2014ApJS..215...15S}.

Continuous advancements in surveys, observational techniques, and the advent of new ground- and space-based telescopes have broadened our understanding of multiple systems and their properties. Much of this progress is summarized in Table 1, which presents the frequency of triple (and higher multiplicity) systems as a function of stellar type/mass, along with the fraction of systems exhibiting evidence for an outer binary with a separation less than 10 AU.

Over the past decade and beyond, \citeauthor{2007ASPC..367..615T} and Borkovits have identified triple and higher-order multiples among low-mass stars and studied some of their other properties, besides their frequency \citep{2007ASPC..367..615T,Tok+14,Tok+17,tok+22}. Zinnecker and later \citeauthor{San+12} have focused on massive stars, demonstrating, as previously mentioned and evident in Table 1, that triples and quadruples are the most prevalent multiplicities among O-stars \citep{San+12,2014ApJS..215...15S}. More recently, in a yet unpublished work, Frost et al. \cite{Fro+25} have leveraged combined data from various observational techniques to reveal that the majority of B-stars also exist in triple- or higher-order systems.

It is crucial to note that the detection of companions is limited by luminosity contrasts. Consequently, the reported fractions correspond to stellar companions with a mass ratio, q, with respect to the primary component, greater than 0.1 ($q>0.1$). Therefore, the detection of companions to higher-mass stars is more limited in range, leading to a decrease in the completeness of multiplicity surveys for higher masses. This suggests that the inferred triple fractions are likely underestimated.

The most comprehensive analysis of the multiplicity census, summarizing a wide range of studies, is presented by \citet{Moe+17}. This work also provided the first detailed characterization of the statistical correlations between different orbital parameters and component masses for binary systems. \cite{Off+23} provides a more recent compilation of multiplicity fractions. However, beyond the overall frequencies of triple systems, current knowledge regarding orbital properties such as mutual inclinations, which are crucial for secular and quasi-secular dynamics discussed below, remains limited. Mass ratios and period ratios, while also not fully characterized, are somewhat better understood, at least for low-mass stars, primarily through studies by \citet{Tok+14,Tok+17}, but are yet missing a complete survey. 

\subsection{Key Orbital and Physical Properties}
Since the complete statistical characterization of orbital properties for triple systems remains elusive, we emphasize that the following summary reflects only the current, partially complete information available for such systems.

\begin{itemize}
    \item The fraction of stars that reside in triple- or higher-order systems exhibits a positive correlation with stellar mass. Among massive O/B-type stars, the fraction of systems in triple or higher order configurations exceeds 50\%, while, for solar-type stars, this fraction is approximately 10$\%$–15$\%$ \citep{Moe+17}. 
    However, an intriguing observation is that for FGK stars, short-period binaries with orbital periods of a few days are highly likely to have an additional stellar companion, forming a triple system. The triple fraction in such systems increases monotonically with decreasing periods, reaching over 95 percent (consistent with 100 percent) for binaries with periods shorter than three days \citep{Tok+06,Ruc+07}.
    \item The prevalence of more compact systems, characterized by outer orbit separations below 10 AU, also increases with stellar mass.
    \item There is tentative evidence suggesting a tendency toward greater orbital alignment in more compact triples, at least for low-mass stars.
    \item Mutual inclinations remain largely unconstrained for most stellar types.
    \item The eccentricities of inner and outer orbits are not well characterized. However, for OB stars, where most systems exist in triple or higher-order multiplicities, the general distribution of binary eccentricities inferred by \citet{Moe+17} likely reflects a combination of inner, more compact binaries and outer, wider orbits.
\end{itemize}

Long-lived, observable triple systems must necessarily be hierarchical and stable (see Section \ref{sec:dynamics} on triple dynamics), which imposes strict constraints on the relationships between the orbital properties of the inner and outer binaries. These constraints arise from dynamical and secular processes. In particular, the closest separation between the outer companion and the inner binary must be at least three times larger than the inner binary's maximum separation (to first order, for comparable masses, $a_{\mathrm{in}}(1-e^{\mathrm{max}}_{\mathrm{out}})/a_{\mathrm{out}}(1-e^{\mathrm{max}}_{\mathrm{in}})\gtrsim 3$; see further details below) to satisfy dynamical stability requirements. The mutual inclinations between the inner and outer binary typically avoid near-perpendicular configurations due to secular/quasi-secular processes that can lead to mergers and collisions in such arrangements (see Section \ref{sec:dynamics} below).

While the phase space of newly formed triple systems may initially include unstable configurations, such systems will rapidly destabilize and/or merge, rendering them unobservable. Consequently, the observed properties of triple systems are inherently sculpted by these processes.
Our knowledge regarding the mutual inclinations and eccentricities of triples remains limited. \citet{Tok+17} finds evidence for non-random alignment between the inner and outer orbits of low-mass triples. A stronger alignment is observed for triples with outer projected separations less than approximately 50 AU, while triples with wide outer orbits exceeding 1000 AU show no alignment with the inner orbits. Additionally, the study finds that orbit alignment decreases with increasing mass of the primary component.

\citet{Bas+24} find that highly compact triple systems (with outer periods shorter than 1000 days) exhibit strong evidence of mutual orbit alignment, as well as a preference for moderate outer eccentricities. They also find that the eccentricity of inner orbits in well-aligned triples tends to be smaller than in non-aligned systems. \citet{tok+19} find that subsystems in compact hierarchies with outer separations less than 100 AU tend to have less eccentric orbits compared to wider hierarchies. 

\begin{table}[h]
\begin{raggedright}
{\scriptsize{}\caption{Multiplicity Statistics of Main-Sequence Stars}
}{\scriptsize\par}
\par\end{raggedright}
{\scriptsize{}}%
\begin{tabular}{lccc}
\hline 
\textbf{\scriptsize{}\cellcolor{blue!10}Survey} & \textbf{\scriptsize{}\cellcolor{blue!10}M$_{1}$} & \textbf{\scriptsize{}\cellcolor{blue!10}Triple+} & \textbf{\scriptsize{}\cellcolor{blue!10}Bin. $<$10 AU}\tabularnewline
\textbf{\scriptsize{}\cellcolor{blue!10}} & \textbf{\scriptsize{}\cellcolor{blue!10}($M_{\odot}$)} & \textbf{\scriptsize{}\cellcolor{blue!10} (\%)} & \textbf{\scriptsize{}\cellcolor{blue!10}(\%)}\tabularnewline
\hline 
{\scriptsize{}\cite{2018MNRAS.479.2702F}} & {\scriptsize{}0.019 \textendash{} 0.058} & {\scriptsize{}$<2$} & {\scriptsize{}$8\pm6$}\tabularnewline
{\scriptsize{}\cite{Bur+07} } & {\scriptsize{}0.05 \textendash{} 0.08} & {\scriptsize{}$0.6\pm0.3$} & {\scriptsize{}\textendash{}}\tabularnewline
{\scriptsize{}\cite{2003ApJ...587..407C}} & {\scriptsize{}0.080 \textendash{} 0.095} & {\scriptsize{}$<3$} & {\scriptsize{}$16\pm6$}\tabularnewline
{\scriptsize{}\cite{2007AJ....133..971A}} & {\scriptsize{}0.06 \textendash{} 0.15} & {\scriptsize{}$<1$} & {\scriptsize{}$14\pm3$}\tabularnewline
{\scriptsize{}\cite{2019AJ....157..216W} 20 pc} & {\scriptsize{}0.075 \textendash{} 0.15} & {\scriptsize{}$2.2\pm1.1$} & {\scriptsize{}$16\pm3$}\tabularnewline
{\scriptsize{}\cite{2019AJ....157..216W} 20 pc} & {\scriptsize{}0.15 \textendash{} 0.30} & {\scriptsize{}$3.6\pm1.0$} & {\scriptsize{}$14\pm2$}\tabularnewline
{\scriptsize{}\cite{2019AJ....157..216W} 20 pc} & {\scriptsize{}0.3 \textendash{} 0.6} & {\scriptsize{}$6.3\pm1.4$} & {\scriptsize{}$15\pm2$}\tabularnewline
{\scriptsize{}\cite{Rag+10} 25 pc} & {\scriptsize{}0.75 \textendash{} 1.25} & {\scriptsize{}$12\pm2$} & {\scriptsize{}$20\pm2$}\tabularnewline
{\scriptsize{}\cite{Tok+14} 67 pc} & {\scriptsize{}0.85 \textendash{} 1.5} & {\scriptsize{}$14\pm2$} & {\scriptsize{}$24\pm2$}\tabularnewline
{\scriptsize{}\cite{2014MNRAS.437.1216D} 75 pc} & {\scriptsize{}1.6 \textendash{} 2.4} & {\scriptsize{}\textendash{}} & {\scriptsize{}$28\pm4$}\tabularnewline
{\scriptsize{}\cite{2021MNRAS.507.3593M}} & {\scriptsize{}1.6 \textendash{} 2.4} & {\scriptsize{}$25\pm5$} & {\scriptsize{}$37\pm5$}\tabularnewline
{\scriptsize{}\cite{Moe+17}} & {\scriptsize{}3 \textendash{} 5} & {\scriptsize{}$36\pm8$} & {\scriptsize{}$46\pm7$}\tabularnewline
{\scriptsize{}\cite{Moe+17}} & {\scriptsize{}5 \textendash{} 8} & {\scriptsize{}$45\pm11$} & {\scriptsize{}$54\pm8$}\tabularnewline
{\scriptsize{}\cite{Fro+25}} & {\scriptsize{}6-15} & {\scriptsize{}$44$} & {\scriptsize{}$>40$}\tabularnewline
{\scriptsize{}\cite{Moe+17}} & {\scriptsize{}8 \textendash{} 17} & {\scriptsize{}$>57\pm15$} & {\scriptsize{}$61\pm10$}\tabularnewline
{\scriptsize{}\cite{San+12,2014ApJS..215...15S}} & {\scriptsize{}17 \textendash{} 50} & {\scriptsize{}$>68\pm18$} & {\scriptsize{}$70\pm11$}\tabularnewline
\hline 
\end{tabular}{\scriptsize\par}
\label{tab:multiplicity}
\end{table}

\section{Dynamics of Triple Systems}
\label{sec:dynamics}

The dynamics of triple stellar systems are rich and complex, encompassing a diverse range of dynamical regimes that govern their long-term evolution. Secular and quasi-secular interactions, in particular, play a critical role in shaping the orbital architecture of hierarchical triples, leading to phenomena such as eccentricity oscillations, inclination oscillations and prograde-retrograde flips, and even chaotic behavior and mergers.

The dynamical evolution of triple systems is dictated by their orbital configurations, which can be broadly classified into distinct regimes exhibiting different types of dynamical behavior. The demarcation between these regimes depends on the strength of the perturbative gravitational effects during the closest dynamical encounters between the constituent stars, the potential cumulative effect of secular/quasi-secular perturbations (operating on timescales longer than the inner orbital period, or both inner and outer orbital periods of a triple), as well as the relative timescale in which the inner and outer binaries interact.

Precisely determining the boundaries between these different regimes remains an active area of research, as discussed below in the context of triple system stability. To first order, and for non-extreme mass ratios, the strength of the perturbative effect distinguishes different dynamical regimes, based on the ratio of the periods or semi-major axes of the outer and inner binaries, or more accurately, by the ratio between the time-weighted innermost approach of the outer orbit to the time-weighted outermost approach of the inner orbit. Additionally, the onset of secular evolution leading to significant orbital changes, as discussed subsequently, strongly depends on the mutual inclination between the inner and outer orbits.

Broadly speaking, triple systems can be categorized into the following dynamical regimes:

1.  \textbf{Highly hierarchical systems:} Characterized by stable inner and outer orbits with non-extreme eccentricities and low mutual inclinations, these systems maintain relatively constant main orbital parameters.

2.  \textbf{Secularly evolving hierarchical systems:} These systems exhibit long-term oscillations in eccentricity and inclination, driven by mechanisms such as the von Ziepel-Lidov-Kozai effect.

3.  \textbf{Quasi-secular systems:} Exhibiting dynamics intermediate between hierarchical and non-hierarchical regimes, these systems often experience extreme eccentricity excitations through semi-period oscillations due to quasi-secular interactions and limited chaotic evolution.

4.  \textbf{Chaotic non-hierarchical systems:} Occurring in non-hierarchical triples, this regime typically leads to the ejection of one component and/or collision/strong interactions between two components.

These regimes, and the transitions between them induced by stellar evolutionary and physical processes, are crucial for comprehending triple systems and their evolution.

\subsection{Hierarchical Systems: $a_{\mathrm{out}}/a_{\mathrm{in}}\gtrsim7.5$; $i\lesssim40^\circ$ or $i\gtrsim140^\circ$}

Hierarchical triples, distinguished by an inner binary and a more distant tertiary companion with low mutual inclination, are inherently stable and exhibit minimal dynamical evolution. Specifically, they maintain approximately constant semi-major axes, eccentricities, and mutual inclination. Unless perturbed by external non-dynamical processes or influenced by stellar evolution, such systems effectively remain static, preserving their overall configurations. 

\subsection{Secularly Evolving Systems: $a_{\mathrm{out}}/a_{\mathrm{in}}\gtrsim7.5$; high inclinations; $40^\circ \lesssim i\lesssim140^\circ$}

Hierarchical triples, characterized by a well-separated inner binary and a more distant tertiary companion with high mutual inclination, are stable and evolve primarily through secular interactions. These interactions induce long-term, gradual changes in the orbital parameters of the system. In this regime, the gravitational perturbations from the outer companion can be averaged over the inner and outer binary orbital periods (so-called double-averaging), leading to a simplified Hamiltonian that describes the long-term evolution of the system.

\begin{figure}
  \centering
        \includegraphics[width=0.6\linewidth]{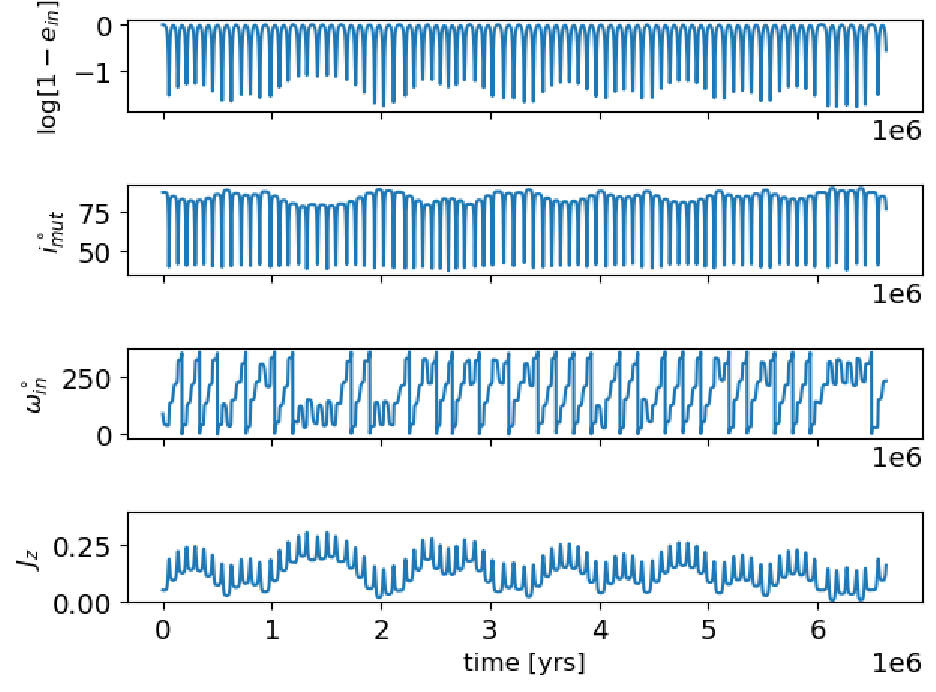}
    \caption{The secular vZLK orbital evolution of a hierarchical triple system. The inner and the outer binaries are well separated, and the system initial parameters are 
     $m_1=m_2=m_3=1$ M$_\odot$; a$_1$=1 AU, $e_1=0.01$, $i_1=87$ degrees, $\omega_{in}=\omega_1=\pi/2$ and $\Omega_{in}=\Omega_1=0$. For the outer binary; a$_2=30$ AU, $e_2=0.0$, $i_2=0$, $\omega_2=\pi/2$ and $\Omega_2=0$; the initial relative mutual inclination $i_{mut}$ is then 87 degrees.
     Quadrupole order effects are hence dominant in this system. We can see that the inner binary undergoes highly regular vZLK oscillations. The numerical simulations are consistent with analytical results for the minimal inclination, shown in dashed black line.}
    \label{fig:vZLK}
\end{figure}

The key dynamical process operating in such hierarchical triples is the von Zeipel-Lidov-Kozai (vZLK) mechanism \citep{vonZeipel1910, Lidov1962, Kozai1962}, which arises when the mutual inclination between the inner and outer orbits exceeds a critical threshold (typically $\gtrsim 40^\circ$). In its classical form, the vZLK mechanism drives large amplitude oscillations in the eccentricity and inclination of the inner binary, with higher eccentricities corresponding to lower inclinations and vice-versa during the oscillations. This process operates when the perturbative effects of the outer companion accumulate adiabatically, and the system evolves secularly on timescales longer than the outer (and inner) orbital periods. An example for the secular vZLK evolution, using N-body simulation (Rebound code) can be seen in Fig. \ref{fig:vZLK}, where we considered a triple system with the same, one Solar mass components and the following initial conditions, for the inner binary  $a_{in}=a_1=1$ AU, $e_1=0.01$, $i_{in}=i_1=87$ degrees, $\omega_{in}=\omega_1=\pi/2$ and $\Omega_{in}=\Omega_1=0$. For the outer binary, $a_{out}=a_2=30$ AU, $e_{\mathrm{out}}=e_2=0.0$, $i_{\mathrm{out}}=i_2=0$, $\omega_{\mathrm{out}}=\omega_2=\pi/2$ and $\Omega_{\mathrm{out}}=\Omega_2=0$; the initial relative mutual inclination $i_{mut}$ is then 87 degrees.
Note that the subscript 1 is used to describe the orbital properties of the inner binary, and 2 is used for the outer binary throughout. Jacobi coordinates are used to characterize the system.

The vZLK mechanism can be elucidated using a Hamiltonian formalism. In this regime, the semi-major axes of both orbits are conserved, while periodic oscillations are introduced in their eccentricities and inclinations. The simplest analytical approach to characterize the dynamics involves expanding the gravitational potential between the inner binary and the outer companion in a multipole expansion and performing orbit averaging over both the inner and outer orbits of the triple \citep[see][for a review]{Naoz2016}. The leading term in this expansion, the quadrupole term, gives rise to the classical vZLK oscillations, while higher-order terms, particularly the octupole term, can induce more complex "eccentric vZLK" behavior when it is non-negligible (see below).

The Hamiltonian for the quadrupole-level vZLK mechanism can be expressed as:

\begin{equation}
H_{\text{quad}} = -\frac{G M_1 M_2}{a_{\text{in}}} \left[ \frac{1}{2} + \frac{3}{8} \frac{a_{\text{in}}^2}{a_{\text{out}}^3 (1-e_{\text{out}}^2)^{3/2}} \left( (1-e_{\text{in}}^2) \left( 3 \cos^2 i - 1 \right) + 5 e_{\text{in}}^2 \sin^2 i \cos 2 \omega_{\text{in}} \right) \right],
\end{equation}

where $G$ is the gravitational constant, $M_1$ and $M_2$ are the masses of the inner binary components, and $\omega_{\text{in}}$ is the argument of periapsis of the inner orbit.

Systems with high mutual inclinations ($40^\circ \lesssim i \lesssim 140^\circ$) evolve through periodic cycles in eccentricity and inclination, which occur on the characteristic vZLK timescale \citep{Ant15,Naoz2016}:

\begin{equation}
\tau^{\mathrm{vZLK}}_{\mathrm{quad}}\sim\frac{16}{15}\frac{a_{2}^{3}(1-e_{2}^{2})^{3/2}\sqrt{m_{1}+m_{2}}}{a_{1}^{3/2}m_{3}G^{1/2}}
=\frac{16}{30\pi}\frac{m_{1}+m_{2}+m_{3}}{m_{3}}\frac{P_{2}^{2}}{P_{1}}(1-e_{2}^{2})^{3/2}
\label{eq:quad_time scale}
\end{equation}


A conserved quantity of this Hamiltonian, is known as the Kozai constant, given by
\begin{equation}
\sqrt{1 - e_{\text{in}}^2} \cos i = \text{constant}.
\label{eq:i_max}
\end{equation}

Hence, the conservation of the Kozai constant enforces that an increase in the eccentricity of the inner binary ($e_{\text{in}}$) during the periodic cycles will be accompanied with a decrease in the mutual inclination, and vice versa. The maximum eccentricity attained, for point particles, is then given by:

\begin{equation}
e_{\text{max}} = \sqrt{1 - \frac{5}{3} \cos^2 i_0},
\label{eq:e_max}
\end{equation}
where $i_0$ is the initial inclination.

In realistic systems with finite-sized stars, the decrease in periastron distance can potentially lead to strong interactions between the inner binary components, significantly affecting the evolution of the stars and the system, potentially even before achieving the maximum possible eccentricity. The driving of high eccentricity and close approaches plays a key role in the evolution of triple systems, leading to strong interactions between the inner binary components. The importance of such processes for the evolution of stellar triples was first recognized by \citet{Har68} and further advanced in \citet{kis+98} and \citet{egg+01}, who focused on the coupled secular and tidal evolution. This was followed and extended to many other dissipative and stellar processes (general relativistic effects and gravitational-wave emissions, mergers, mass transfer, etc.) by numerous studies over the last two decades, as discussed below.

These pioneering works and subsequent extensive studies have led to our current understanding of vZLK evolution and its eccentric vZLK and quasi-secular related processes and extensions (see below) as key processes in the evolution of gravitating triple systems across all scales, particularly in triple stellar systems.

\subsubsection{Octupole-Level Effects and the Eccentric vZLK evolution}
While the quadrupole approximation provides a good description of the vZLK mechanism in many cases, higher-order terms in the multipole expansion can become important, especially when the outer orbit is eccentric and when the masses of the three bodies are significantly different. Quantitatively defined, the importance of the octupole term is given by
\begin{equation}
\epsilon_{oct}=\frac{m_{1}-m_{2}}{m_{1}+m_{2}}\frac{a_{1}}{a_{2}}\frac{e_{2}}{1-e_{2}^{2}},
\end{equation}
and therefore, as the mass of the tertiary companion increases significantly, the eccentric vZLK mechanism transitions back to the regular vZLK regime, where the influence of the outer companion becomes more akin to that of a distant, circular perturber, as in the case of a binary orbiting a massive black hole, or binary asteroids orbiting the Sun.

The octupole-level derivation and basic characterization have been first explored by \citet{Har68} and later by \citet{For+00} and \citet{Bla+02}. A major advance in understanding its key importance, and providing detailed study and extensions have been initiated by \citeauthor{Nao+11} and collaborators \citep{Nao+11,Nao+13}; and studies by others \citep[e.g.][and references therein]{Kat+11,Kle+24}. In this regime, termed the Eccentric vZLK regime, the secular evolution can become more acute, including more extreme eccentricity oscillations than in the quadrupole regime and even flips (prograde-retrograde) in the orientation of the inner orbit, and the evolution becomes more chaotic.
Given the more cumbersome and more complex nature of this regime, we refer to Naoz's review \citep{Naoz2016} for detailed derivation and discussion; key issues are the dependence on the outer eccentricity of the triple and the relation between the triple component masses. For equal mass cases, for example, the octupole term vanishes. 

The timescale corresponding to the octupole level perturbations is
\begin{equation}
\tau_{\text{vZLK-oct}}\sim2\pi\frac{a_{\text{out}}^{4}\left(1-e_{\text{out}}^{2}\right)^{5/2}\left(1-e_{\text{in}}^{2}\right)^{1/2}\left(m_{1}+m_{2}\right)^{3/2}}{G^{1/2}m_{3}\left|m_{1}-m_{2}\right|e_{\text{out}}a_{\text{in}}^{5/2}}.\label{eq:octupole_time}
\end{equation}
The long-term modulation of vZLK cycles occurs on this
timescale, which is longer that the vZLK timescale.

\begin{figure}
  \centering
        \includegraphics[width=0.6\linewidth]{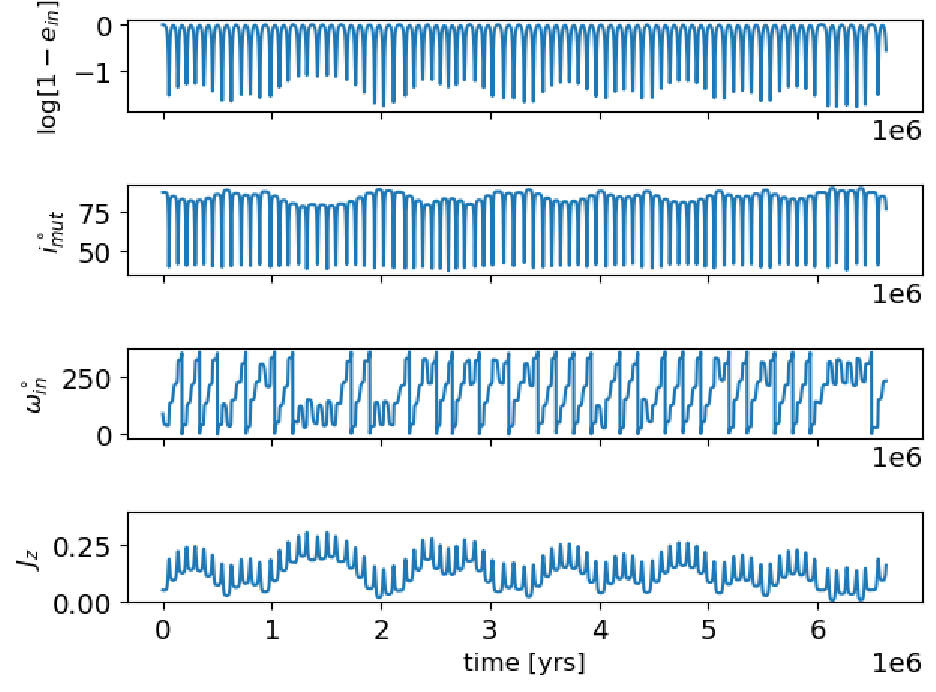}
    \caption{The orbital evolution of a triple systems in the Octupole-level regime. The system initial parameters are $m_1=1.2$ M$_\odot$, $m_2=0.1$  M$_\odot$ and $m_3=0.1$ M$_\odot$. The initial conditions for the inner orbit are  $a_1=1$ AU, $e_1=0.01$, $i_1=87$ degrees, $\omega_1=\pi/2$ and $\Omega_1=0$; and for the outer orbit $a_2=30$ AU, $e_2=0.7$, $i_2=0$, $\omega_2=\pi/2$ and $\Omega_2=0$; the initial relative mutual inclination is then 87 degrees.  The outer orbit is eccentric in this simulation, and non-equal, low mass-ratio is considered. Note the less-ordered behavior and modulations of the vZLK oscillations, showing differing eccentricity peaks, compared with the quadrupole-regime case in Fig. \ref{fig:vZLK}.}
    \label{fig:ecc_vZLK}
\end{figure}

When considering the octupole term, in cases it becomes non-negligible, it introduces additional complexities to the dynamics, introducing terms that depend on the eccentricities of both the inner and outer orbits, as well as the argument of periapsis of the outer orbit ($\omega_{\text{out}}$). This leads to a richer dynamical behavior and can drive triple systems to reach a wider range of extreme orbital configurations, inaccessible in the classical vZLK regimes, with potentially stronger inner-binary interactions, and a wider range of mutual inclination evolution, and potentially even flipping between prograde and retrograde mutual inclinations (in the case of a test particle limit in the inner binary). Given that the phase space for the conditions for the eccentric vZLK to become important for stellar triples is quite large, it plays a major role in the evolution of a wide variety of triple stellar systems. 
Fig. \ref{fig:ecc_vZLK} shows such example. In this case we consider more unequal masses, $m_1=1.2$ M$_\odot$, $m_2=0.1$  M$_\odot$ and $m_3=0.1$ M$_\odot$. The initial conditions for the inner orbit  are $a_1=1$ AU, $e_1=0.01$, $i_1=87$ degrees, $\omega_1=\pi/2$ and $\Omega_1=0$; and for the outer orbit $a_2=30$ AU, $e_2=0.7$, $i_2=0$, $\omega_2=\pi/2$ and $\Omega_2=0$. The octupole parameter is then $\epsilon=0.038$ (while the examples in Figs. \ref{fig:vZLK} and \ref{fig:quasi_vZLK} have $\epsilon_{vZLK}=0$).

\subsubsection{Inverse Kozai Mechanism}
A further extension of the vZLK mechanism is the \textit{inverse-Kozai} case, where the inner binary is far more massive than the outer companion, leading to a different dynamical evolution. In such systems, the roles of the inner and outer orbits are effectively reversed, and the inner binary exerts a significant perturbation on the outer orbit. This leads to oscillations in the outer orbit's eccentricity and inclination, somewhat analogous to the classical vZLK mechanism, but with the perturbations originating from the inner binary \citep[e.g.][]{Nao+17,Vin+18}.

\subsection{Quasi-Secular Regime; $3\lesssim a_{\mathrm{out}}/a_{\mathrm{in}}\lesssim7.5$; high mutual inclinations}
The vZLK mechanism operates under the assumption of adiabatic secular evolution of the inner and outer orbits of the triple. However, in systems where the perturbation of the inner orbit by the outer companion becomes significant on dynamical timescales, a different regime emerges. This typically occurs when the pericenter approach timescale becomes comparable to the inner binary orbital period, leading to interactions that are faster than those in the fully secular regime but still governed by long-term orbital evolution. This is known as the quasi-secular regime, or single averaged regime (since orbit-averaging over the outer orbit, is no-longer consistent). 

In this regime, whose importance was first highlighted and analyzed in \citet{Ant+12}, the inner and outer orbits exchange angular momentum and energy more rapidly. This can lead to extreme oscillations in eccentricity and inclination over shorter timescales than in the secular regime. Systems in the quasi-secular regime are often prone to chaotic behavior, with eccentricities reaching unity even for lower mutual inclinations, and no longer completely adhering to a strict low-inclination–high-eccentricity relation.

In the quasi-secular regime, the averaging techniques used to derive the secular Hamiltonian are not strictly valid, as the inner and outer orbits exchange angular momentum and energy more rapidly. This regime and its key importance were first explored through few-body simulations by \citet{Ant+12}, followed by few-body modeling over a wide range of conditions and stellar populations \citep{Kat+12,Hai+18}. The more extreme and chaotic behavior allows for extreme and rapid eccentricity excitations, potentially leading to high-eccentricity mergers and collisions that would otherwise be inaccessible in the secular regimes \citep{Ant+12,Kat+12,Ant+14}. More analytical approaches have been developed to explore this regime, both through single-averaged Hamiltonians (not averaging over the outer period) and perturbative approaches to double-averaged Hamiltonians \citep{Luo+16,Hai+18,Gri+18,Kle+24}, and the recent identification of the Brown-Hamiltonian to analyze the evolution in these (and possibly all secular) regimes \citep{Tre23,Gri+24}.

\begin{figure}
  \centering
        \includegraphics[width=0.6\linewidth]{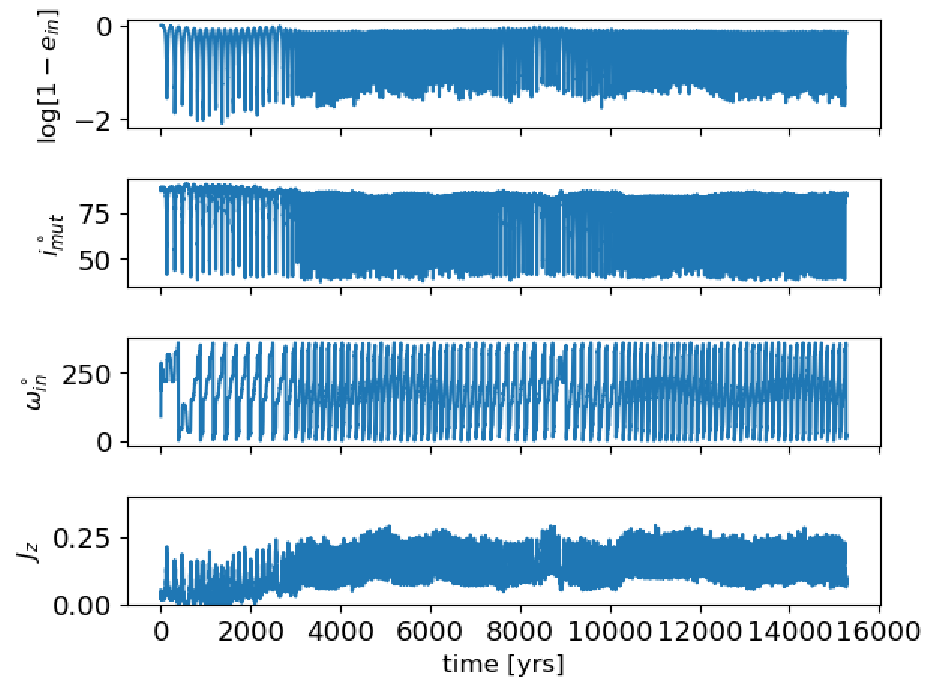}
    \caption{The Orbital evolution of a triple system in the quasi-secular regime. The outer companion is now close to the inner binary, with  $a_1=1$ AU, $e_1=0.01$, $i_1=87$ degrees, $\omega_1=\pi/2$ and $\Omega_1=0$ for the inner binary, and 
$a_2=6$ AU, $e_2=0.4$, $i_2=0$, $\omega_2=\pi/2$ and $\Omega_2=0$, for the outer binary (i.e. with a mutual initial inclination of 87 degrees). As can be seen, a much more irregular behavior is observed, showing eccentricity peaks changing chaotically with time. In this case the masses are equal and the systems has no contribution from the the octupole regime}
    \label{fig:quasi_vZLK}
\end{figure}

In Fig. \ref{fig:quasi_vZLK} we show an example for the evolution of a triple system with all components having the same, one Solar mass components, m$_1$=m$_2$=m$_3$ (and hence no contribution from the octupole regime) in the quasi-secular regime. Its initial conditions are given by 
$a_1=1$ AU, $e_1=0.01$, $i_1=87$ degrees, $\omega_1=\pi/2$ and $\Omega_1=0$
for the inner binary, and 
$a_2=6$ AU, $e_2=0.4$, $i_2=0$, $\omega_2=\pi/2$ and $\Omega_2=0$, for the outer binary. As can be seen, a much more irregular behavior is observed, showing eccentricity peaks changing chaotically with time. In this case the masses are equal and the systems has no contribution from the the octupole regime.   

The conditions for this regime to become important require more compact triple configurations than the vZLK/eccentric-vZLK regimes but still apply to a wide range of triple stellar systems. Indeed, \cite{Luo+16} identified the level of importance of the quasi-secular (single-average; SA) effects by
 $\epsilon_{SA}$, which  describes the strength of single averaged/quasi-secular terms:
 \begin{equation}
 \epsilon_{SA}={(\frac{a_{1}}{a_{2}})}^{3/2}\frac{1}{(1-e_{2}^{2})^{3/2}}\frac{m_{3}}{((m_{1}+m_{2})m_{1})^{1/2}}
 \end{equation}

For example, the systems show in Figs. \ref{fig:vZLK} and \ref{fig:ecc_vZLK} have small $\epsilon_{\text{SA}}$ with $\epsilon^{\text{vZLK}}_{\text{SA}}\simeq0.004$, and  $\epsilon^{\text{ecc-vZLK}}_{\text{SA}}\simeq0.001$ for the vZLK and eccentric-vZLK regimes, respectively, with the quasi-secular evolution depicted in Fig. \ref{fig:quasi_vZLK} having $\epsilon^\text{quasi}_\text{SA}\simeq0.04$.

The quasi-secular regime allows for more extreme configurations even compared to the eccentric vZLK processes and may operate even for non-eccentric outer orbits and/or extreme mass ratios, such as binary stars orbiting a massive black hole.

\subsection{Relativistic effects on secular evolution}

Relativistic effects can significantly impact the secular evolution of hierarchical triple systems, particularly by influencing secular/quasi-secular processes. The dominant relativistic effect is the precession of the argument of pericenter due to General Relativity (GR). This precession can suppress vZLK oscillations by disrupting the coherent evolution of the argument of pericenter, which is essential for driving the vZLK cycles. Specifically, the presence of GR corrections can reduce the region in parameter space where large eccentricities (or inclinations) can be achieved \citet[see][for a review]{Naoz2016}.

In certain parameter regimes, GR effects can lead to resonant eccentricity excitation. This phenomenon arises when the timescales of GR precession and vZLK oscillations become comparable, resulting in a resonant interaction that can enhance the eccentricity of the inner orbit \citep{Nao+13b}.

Further studies by \citet{wil17} and \citet{Lim+20} explored the effects of including higher-order terms and different mass ratios, revealing even richer evolutionary pathways, including more extreme behaviors in eccentricity and inclination changes.

\subsection{Non-hierarchical unstable triples -- the chaotic regime: $a_{\mathrm{out}}/a_{\mathrm{in}}\lesssim3$}

Non-hierarchical triple systems, where the three stars have comparable separations and orbital periods, are characterized by chaotic dynamics. In this regime, the gravitational interactions between the three stars are strong and, for all practical purposes, unpredictable. This leads to complex and unstable orbital evolution. The study of such chaotic behavior in triple systems, pioneered by Poincaré, laid the foundation for the field of chaos theory.

A key indicator of chaotic dynamics is the Lyapunov exponent, which quantifies the rate at which nearby trajectories in phase space diverge. A positive Lyapunov exponent signifies chaotic behavior, where small perturbations can lead to drastically different outcomes. This sensitivity to initial conditions is a hallmark of triple systems in this regime. While specific stable periodic configurations can be mathematically derived within this regime, they occupy a negligible portion of the phase space and are not relevant to realistic stellar systems.

In this regime, the strong gravitational perturbations among the three stars induce significant changes in their orbits on short, dynamical timescales. This results in a chaotic evolution often characterized by repeated ejections of one component to a large distance, temporarily forming a hierarchical configuration. This ejected star eventually returns to the inner binary, leading to a close encounter and a chaotic interaction ("scramble"), often resulting in another temporary ejection. This process typically culminates in the complete ejection of one of the stars from the system, leaving behind a stable binary. While the precise evolution of any individual chaotic system is unpredictable, the statistical distribution of outcomes for an ensemble of such systems can be characterized.

For stars with finite sizes and/or dissipative interactions, such as tides or collisions, additional outcomes are possible. These include stellar collisions and mergers during close encounters. Such interactions can lead to the formation of a merger product and a binary companion \citep{Dav+94}, or even a new, stable hierarchical triple system if a tidal capture occurs, rapidly shrinking and circularizing the inner binary.

\subsubsection{Evolution of unstable triples and binary-single encounters}

The evolution of non-hierarchical triple systems cannot be modeled using secular or quasi-secular approaches and is therefore explored numerically through few-body simulations. Such simulations have been performed since the 1960s, with progressively larger samples and improved characterization of the outcomes \citep[e.g.][]{age+67,Sta+72,Hil75,Hut+83,ano86}. These simulations were used to probabilistically characterize the final outcomes of such encounters and the associated timescales.

These numerical studies, augmented by analytical analyses, have identified general trends in the evolution of unstable triples. In particular, \citet{Heg75,Heg+93} and \citet{Hil75} identified what is known as Heggie's law: \textbf{Hard binaries get harder, soft binaries get softer}. In a stellar encounter, a binary star system with internal energy $E_{\text{bin}}$ interacting with a perturber of mass $m_3$ incoming with a relative velocity $v_{\text{rel}}$ is termed "hard" if $E_{\text{bin}} > m_3 v_{\text{rel}}^2$. The surviving binary in such an encounter will tend to become harder (i.e., have increased binding energy compared to the original binary, which for equal-mass components generally translates to a more compact remnant binary). Conversely, a "soft" binary for which $E_{\text{bin}} < m_3 v_{\text{rel}}^2$ will tend to "soften" (i.e., decrease the binding energy of the remnant binary).

In such encounters, or generally, in a chaotic triple, the least massive component is most likely to be ejected, with the remnant binary being more massive and/or more compact. The accumulation of many encounters will therefore lead to progressively harder binaries becoming even more compact and/or having more massive components, while soft binaries become softer, with lower binding energy, and eventually become unbound and "evaporate," releasing two single stars. Soft binaries are therefore generally short-lived, while hard binaries survive. The ejected star in encounters with hard binaries gains kinetic energy. Consequently, hard binaries serve as a kinetic heating source in dense environments where binary-single encounters are frequent, such as stellar clusters, particularly dense globular and nuclear clusters.

Harder, more compact binaries are more likely to become interacting binaries, exhibiting diverse phenomenology such as mass transfer, tidal interactions, common-envelope evolution, and gravitational-wave inspiral. Indeed, globular clusters are observed to have a significantly increased frequency of interacting binaries and merger products, such as X-ray binaries, recycled pulsars, cataclysmic variables, and blue stragglers. They are also suggested to play a key role in producing gravitational-wave mergers. Therefore, the dynamical evolution of chaotic triples arising from binary-single encounters is one of the main drivers affecting the evolution of stellar clusters and their constituent stellar populations and interacting binaries.

While binary hardening is a key aspect of binary-single encounters and the evolution of binary populations in clusters, this aspect only focuses on the binary's \emph{energy}. The resulting binaries from disrupted unstable triples and binary-single encounters also change their angular momentum. In particular, while hardening in the energy phase space leads to more compact binaries that could eventually strongly interact, the change in angular momentum can be far more effective in driving strong interactions. Repeating encounters allow remnant binaries to explore a wide range of angular momentum, producing binaries on highly eccentric orbits, where the pericenter approach is sufficiently small to trigger strong binary interactions, potentially even for initially very wide binaries. Hence, strong binary interactions are driven more by angular momentum changes than by binary hardening. While this was already pointed out and shown by Hills early on \citep{Hil84}, it received very little attention compared with Heggie's law, while its implications might be sometimes more important.

Another aspect not fully characterized in such approaches is the focus on the overall dynamical encounter outcome for three-point-mass particles. During the chaotic evolution, stars, which are finite-size objects, can strongly interact through tidal interactions, collisions, or gravitational-wave emission. Such interactions dissipate energy from the system and would change the encounter outcomes, potentially even leading to stable triples or merger product binaries, as mentioned above. Accounting for the complex evolution during an encounter, and not only the final outcome of a disrupted triple, is therefore of key importance but is generally more complex to model, depending on the type of dissipative process. Numerical studies accounting for dissipative aspects have shown such effects change the type of gravitational-wave mergers produced in encounters of compact objects, showing the production of eccentric mergers, a wide variety of collision products, and overall different outcomes compared with simplified simulations that neglected these various interactions \citep{Mcm86,Dav+94,Sig+95,Lom+03,Fre+04,Gul+06,Sam+14,Sam+17}.

While extensive N-body simulations can be used to provide a probability distribution of outcomes, each different set of initial conditions requires new (and computationally costly) simulations. However, the study of chaotic triples has been revolutionized over the last few years with the use of statistical physics methods, allowing for analytical/simple numerical characterization of the probabilities and branching ratios for the chaotic evolution outcome. Building on the foundations laid by \citeauthor{Mon76} \citep[in energy phase space only]{Mon76,Mon76b}, \citet{Sto+19} and \citet{gin+21} derived a complete statistical solution for non-hierarchical triples by assuming ergodicity, meaning that the system explores all accessible phase space. This allows for closed-form distributions of outcomes (e.g., binary orbital elements) given conserved quantities like energy and angular momentum. They found good agreement with numerical simulations, particularly for "resonant" encounters exhibiting chaotic evolution. \citet{Kol21} has suggested an alternative flux-based approach, which, while not yet providing full orbital solution distributions, may provide more accurate results for the branching ratios of ejected masses in multi-mass encounters.

\citet{gin+21} further extended the phase-space approach to analyze not only the final outcome but also the intermediate states, modeling the interaction as a random walk between hierarchical phases punctuated by close triple approaches. They also showed that a detailed-balance approach can be used to produce the same results. Their solution provides probability distributions for the remnant binary's orbital parameters. Importantly, this framework allows for the inclusion of dissipative forces like tides or gravitational-wave emission, which are often challenging to incorporate into direct few-body integrations, and aspects of collisions during encounters. These can also be used to account for external potential effects and to provide detailed probability distributions arising from the cumulative effects of many encounters \citep{Gin+21b,Gin+23}.

These statistical solutions provide powerful tools for understanding the outcomes of three-body interactions in stellar systems, with implications for binary evolution, star cluster dynamics, and the formation of compact objects.

\subsubsection{Triples stability criteria}

The stability of triple stellar systems, particularly hierarchical triples, has been a subject of extensive study, both analytically and through numerical simulations. Here, we define stability in the sense that all three stars remain bound to the system (i.e., none are ejected), and the hierarchy of the inner and outer binaries is maintained over time (i.e., no exchanges occur). Dynamical evolution may also lead to a direct collision/merger between two or more components, which would also terminate the existence of a triple. However, here we first consider the stability of point-mass triples, neglecting the effects of finite size and collisions. While early studies referred to some systems as unstable if their eccentricity and inclination evolved, as discussed above, such secular evolution is expected for a wide range of hierarchical triple systems, and these are considered here as stable triples.

A key factor determining the stability of triple systems is the strength of the perturbation of the third companion on the inner binary. This, in turn, depends on the separation between the inner binary and the outer tertiary companion, as well as the masses of the triple components. For a triple system to be dynamically stable over long timescales, the outer orbit must be sufficiently wide to prevent gravitational perturbations from the outer companion from destabilizing the inner binary.

\citet{Evans68} was among the first to discuss the hierarchy of multiple systems. Early studies by \citet{Har72a} investigated the general three-body problem numerically and derived some of the first stability criteria for such systems. In particular, it was found that for a triple to be stable (for the equal-mass case), the closest approach (pericenter, $q$) of the outer orbit to the inner orbit needs to be larger than 3.5 times the semi-major axis of the inner orbit ($q_2/a_1 = a_2(1-e_2)/a_1 > 3.5$). It was also found that retrograde orbits are more stable than prograde ones, requiring a minimal ratio of only 2.75 (while noting that even closer orbits might exist under some conditions). Similar analytical results, generally consistent with the numerical results, were found by various authors \citep{Boz76,Mar+75,Sze+77} and also provided a mass dependence. However, they did not recover the higher stability of retrograde orbits and, in fact, predicted greater stability for prograde orbits, in contrast with the numerical results.

\citeauthor{Har72a} also noticed quasi-instability near perpendicular configurations, by which he meant that such configurations lead to collisions of the inner binary components rather than the unbinding of the system, i.e., not the type of instability directly discussed here. Indeed, this is to be expected from the secular vZLK process, which leads to radial orbits for initially perpendicular inner and outer orbits. 

Very few analytical studies were done for the case of general inclinations. However, \citet{Ina80} showed analytically that systems should become progressively and monotonically more stable with higher inclinations, consistent with retrograde orbits being more stable. Nevertheless, numerical results showed that the increase in stability with inclination is not completely monotonic; intermediate inclinations exhibit a dip in the stability increase.

\citet{per+09a} and \citet{Gri+17} have shown that secular vZLK/quasi-secular-like evolution affect the long-term stability of triple systems. While previous studies explored the initial eccentricity of the systems, they did not account for the secular eccentricity evolution. Hence, one needs to consider the expected minimal approach that occurs on secular timescales when the amplitude of eccentricity oscillations is maximized. Accounting for the secular evolution reproduced the non-monotonic inclination dependence, with the dip occurring in the inclination regime where such secular/quasi-secular processes are effective. While this was initially shown for cases with very low mass ratios (a low-mass binary system orbiting a massive object), the same behavior was shown to hold for stellar triples with comparable masses \citep{Myl+18,Vyn+22}.

\citet{Mar08} developed a different, overlapping resonances approach to study triple stability and derived an algorithm to derive the stability criteria for coplanar systems. However, a generalized version for arbitrary inclinations following this method has not been published.

Over the last two decades, the most widely used stability criteria for triples have followed the combined analytical and numerically derived fit by \citet{Mar+01}, based on their analytical derivation of stability criteria for coplanar prograde configurations, augmented with a simplified inclination dependence based on fits to few-body simulations. While generally providing a good fit, it does not include any dependence on the inner binary eccentricity, and the inclination dependence is quite inaccurate (although derived from fits to few-body simulations, these were not followed for more than 100 orbits and likely did not capture the full effects of secular evolution). A better, more accurate numerical combination of analytical derivation and numerical fit to extensive few-body simulations was later found by \citet{Myl+18}, which provided a more complex inclination dependence fitted to the numerical findings.

More recently, \citet{Vyn+22} have followed the results by \citet{Gri+17} to provide analytical stability criteria that also account for the expected inclination dependence affected by vZLK secular processes, though not using the higher-order corrections for quasi-secular evolution discussed by \citet{Gri+17}. This criterion provides an excellent fit for the simulation results. Nevertheless, they have also used machine learning methods trained on a large set of few-body simulations and provided a stability-criteria numerical function with an even more precise fit to the numerical few-body simulation results. We believe the current best numerical, analytical, and machine learning-based results are those obtained in these recent studies \citep{Myl+18,Vyn+22}.

The timescale for an unstable system to disrupt was studied in few-body simulations (e.g. \citep{ano86}), and discussed by \citeauthor{val+06} \citep{val+06}. However, very few analytical studies have explored this issue. \citet{Mus+20} made use of a random-walk approach to derive such a distribution. Though not directly applicable to comparable-mass triples; \citet{Zha+23} also studied this problem.

\section{Stellar Evolution and Dissipative Processes}

The dynamical evolution of triple stellar systems is deeply intertwined with the internal evolution of the stars themselves. As stars evolve, they undergo processes such as mass loss, and mass transfer, and change their physical properties (radii, density profile), which can directly impact the orbital dynamics of their companion stars. In addition, dissipative processes, including tidal interactions and common-envelope evolution, can significantly alter the configuration and stability of triple systems. This section explores the main stellar evolutionary processes and the role of dissipation in shaping the outcomes of triple systems.

\subsection{Eccentricity excitation by secular and quasi-secular processes}

As discussed above, secular and quasi-secular processes in triple systems can excite the eccentricities of the inner binary, driving the stars to a close pericenter approach. \citet{Har68} was among the first to recognize that secular processes in triples could lead to strong interactions and even mergers of the inner binary components. The various combinations of secular and quasi-secular processes with dissipative effects are therefore a key aspect of triple stellar evolution.

In general, eccentricity excitation by secular processes leads to progressively closer pericenter approaches, to the point where close dissipative interactions affect the evolution, most typically by extracting energy and angular momentum, thereby shrinking and circularizing the orbit. Such processes typically introduce precession of the orbit, which may quench the secular evolution at some point. Eccentric vZLK, and to a much greater extent, quasi-secular evolution, can drive much faster eccentricity excitations, such that the inner binary may be excited to high eccentricity on a dynamical timescale. Consequently, the binary may experience a sudden jump into a strong interaction phase, potentially even a direct collision, without being previously affected by gradual dissipative processes.

In the following, we briefly discuss various channels for coupled secular/quasi-secular evolution and dissipative processes.

\textbf{Tidal interactions and magnetic braking:} \citet{Maz+79} suggested that coupled secular evolution and tidal friction can drive inner binaries to shorter periods. While they did not discuss vZLK evolution, but rather small-amplitude secular evolution (that can happen even at low mutual inclinations) they introduced this possibility. \citet{kis+98} and \citet{egg+01} showed that vZLK evolution drives close pericenter approaches, which can then induce tidal interactions between the inner binary components. In turn, the tidal dissipative evolution could lead to the shrinkage of the binary orbit. Therefore, the combined action of vZLK cycles and tidal friction can drive the shrinkage of binary orbits and the formation of short-period binaries. Interestingly, it is observed that very short-period FGK binaries are likely to have a third companion \citep{Tok+06}, as discussed above, which might be related to this process, or a similar process during the pre-main-sequence evolution of triples \citep{moe+18}. However, this might also be an unrelated result of the star formation processes that produce triple systems.

Since the highest eccentricities correspond to the lowest inclinations during vZLK cycles, shrinking binaries may have a preference towards lower inclinations (close to $40^\circ$ or $140^\circ$), as was suggested to be the case for similarly shrinking planetary systems \citep{Fab+07} and binary asteroids \citep{per+09a}. This may even lead to binary mergers and the production of rejuvenated blue straggler stars and might explain their existence in some wide binary systems \citep{per+09b}. The eccentric-vZLK process can drive even stronger interactions and further increase the production of shrinking binaries and merger products \citep{Nao+14}, and quasi-secular effects may further introduce more interactions (an effect not yet studied in depth in this context).

Similar evolution can affect various types of stars and compact objects in triples and therefore lead to the production of a wide variety of compact interacting binaries, which can then also be affected by mass transfer \citep{Maz+77,Mak+09,Nao+16,Sha+24}, as we further discuss below. In principle, magnetic braking can also introduce dissipation and binary orbital shrinkage and can therefore introduce similar effects as the coupling of tidal evolution with secular/quasi-secular dynamics \citep[e.g.][]{Egg06,Egg+06,Bat+18}.

\textbf{GW emission and inspirals:} For compact object inner binaries in triples, for which tides might be far less effective, a sufficiently close approach can introduce large dissipation through gravitational-wave (GW) emission. Similarly to the combined secular evolution and tidal friction, combined secular evolution and GW emission can therefore drive increased rates of compact binary mergers in triples as suggested in \citet{Tho+11}. In \citet{Ant+12}, it was shown that rapid secular, and even more so quasi-secular, evolution can lead to sufficiently high eccentricities and not only induce such mergers but drive them to merge while still eccentric, not allowing for the GW dissipation to circularize them before the final merger \citep[see also][]{Ant+14}. The combined eccentric-vZLK/quasi-secular evolution of compact objects in triples therefore drives fast and potentially eccentric GW mergers.

\textbf{Eccentric mass-transfer and common-envelope evolution:} Close interactions between stars may result in mass transfer or common-envelope evolution. In isolated binary systems, the evolution towards mass transfer and common-envelope evolution typically occurs gradually, either due to the slow stellar evolution that stars experience as they evolve off the main sequence to become giants, thereby slowly expanding to fill their Roche lobe, or due to slow tidal evolution, even on the main sequence, that drives orbital shrinkage. Such slow evolution may allow for the circularization of the binary orbit, such that most of the mass-transfer/common-envelope evolution occurs while the binaries have low or even zero eccentricities. When driven by secular processes, fast eccentricity excitation can introduce mass transfer and common-envelope interactions while the inner binaries in a triple are eccentric, and even highly eccentric. Therefore, while eccentric mass transfer/common-envelope may occur to some extent even for isolated binaries, such eccentric mass transfer becomes a potentially key aspect of triple evolution. The evolution of such eccentric binaries through this evolutionary stage could significantly differ from that of circular binaries. 

While hardly explored in the triple context, mass transfer in eccentric binaries has been studied by \citet{Reg+05}, \citet{Sep+07}, \citet{Sep+09}, \citet{Sep+10}, \citet{Laj+11}, \citet{Dav+13}, and \citet{Ham+19} through numerical and analytical studies. An eccentric mass-transfer coupling to triple secular evolution has been introduced into a population synthesis code by \citet{Ham+21}. Eccentric common-envelope evolution has been little studied; eccentric pre-common-envelope conditions were studied in \citet{vic+21}, while hydrodynamical simulations of the eccentric common envelope have been studied in \citet{Gla+21}. This combined secular/quasi-secular evolution is therefore still little understood and requires further studies.

\subsection{Mass Loss and Mass Transfer in Triple Stellar Systems}

Mass loss and mass transfer processes are critical in shaping the dynamical evolution of triple stellar systems. While the secular/quasi-secular driving of close interactions and eccentric mass transfer have been discussed above, here we focus on the effects of mass loss on triple dynamics and mass-transfer processes unique to triple systems. Such processes drive mass and orbital changes, influence secular interactions, and can even induce dynamical instabilities, ultimately influencing the long-term fate of triple systems.

The type of effects depends on the mass-loss timescale compared with the timescale for dynamical and secular evolution, and on which a triple component is losing/transferring mass. Below, we first discuss types of mass transfer in triples, from the outer companion to the inner binary, and from an inner binary component. We then discuss triple hierarchy change due to two main cases of mass loss: adiabatic mass loss, occurring on timescales much longer than the dynamical (orbital period) timescales of the system, and prompt mass loss, occurring on timescales shorter than the dynamical timescales.

\subsubsection{Circumbinary and circumstellar mass transfer and common envelope evolution}

In triples, mass transfer can occur in three distinct ways: (1) from one of the inner binary components to its close companion; (2) from the inner binary to the third, outer companion (circumbinary mass transfer); or (3) from the outer companion to the inner binary (circumstellar triple mass transfer).

Circumstellar mass transfer from an outer companion onto an inner binary has been shown to give rise to accretion onto both stars of the inner binary, and some specific cases were studied analytically and in hydrodynamical simulations \citep{Sok04,deV+14,Por+19,Gla+21}. It was suggested that such mass transfer may give rise to a circumbinary accretion disk \citep{Lei+20} or a common envelope engulfing the inner binary \citep{Ibe+99,Gla+21}. This common envelope may lead to the merger of the inner binary or otherwise affect its orbit \citep{Ibe+99,Lei+20,Dis+20,Gla+21,Dor+24, Kum+24,Val+24}, or to mass accretion that could rejuvenate the inner binary components \citep{Por+19,Ibe+99b} or produce cataclysmic variables \citep{Ibe+99b}. 

Circumbinary mass transfer from the inner binary to the third, outer companion has not been explored in detail, to the best of our knowledge. In \citet{Egg+86}, the merger product of an inner binary was suggested to expand and engulf a third companion, leading to a common-envelope evolution. In general, mass transfer from an inner binary to a third companion can occur. However, the stability requirements of triple systems suggest that the third companion would typically not be very close, allowing for less accretion. Nevertheless, wind Roche-lobe overflow may still occur even in wider triples, and some triples may be sufficiently compact to allow for direct mass transfer or even engulfment of the third companion and an effective triple common envelope. These aspects require further investigation.

Mass transfer solely between the inner binary components can alter the triple configuration, as we discuss below. In particular, it can lead to hierarchy changes even when the mass loss is conservative, and no mass is lost from the triple system as a whole.

\textbf{Common-envelope evolution:} Common-envelope (CE) evolution is a critical phase in the evolution of close binaries and plays a crucial role in the formation of compact object binaries. In a triple system, common-envelope evolution can occur either through the circumstellar or circumbinary cases discussed above. During this phase, the drag forces exerted by the envelope cause the stars in the inner binary to spiral inward, leading to a dramatic reduction in their orbital separation \citep{Gla+21}. In the circumbinary case, the third companion can be affected by the drag, but the mass loss from the inner binary also expands the outer orbit, making the final outcome difficult to predict.

Therefore, the outcomes of common-envelope evolution in triple systems are likely quite diverse. In some cases, the inner binary emerges as a close compact object binary, orbited by the core of the third companion. In others, the inner binary merges to form a binary with the core of the third companion. In yet other cases, the binary might inspiral close to the third companion's core, and they will become an unstable triple and be disrupted, ejecting one of the stars/cores \citep{Sab+15,Gla+21,Mor+23}. Such outcomes can also affect the physical properties of the remnant stars/compact objects \citep{sok22}.

\subsubsection{Hierarchy change due to mass loss and mass transfer in triple stellar systems}

\textbf{Impact of adiabatic mass loss on orbital evolution:} As stars evolve off the main sequence, they experience significant mass loss through stellar winds (massive stars can have non-negligible mass loss even during their main-sequence evolution), especially during the asymptotic giant branch (AGB) phase. Adiabatic mass loss due to such stellar winds leads to changes in the system masses, which in turn affect the orbital periods and separations. Such effects have little dependence on the specific phase of the orbit, and the evolution can be thought of as orbit-averaged due to the slow adiabatic evolution. Such processes were mainly introduced and studied using few-body simulations by \citet{per+12} and \citet{Sha+13}, and through a secular code, by introducing mass-loss evolution into the secular evolution of the Hamiltonian \citep{Mic+14}. These were later incorporated into population synthesis models \citep[e.g.][]{Ham+22b,Toonen2022}.

While the inner and outer binary eccentricities are approximately conserved in adiabatic mass loss, the semi-major axis of a binary changes. The fractional change in the orbital semi-major axis ($\Delta a/a$) can be expressed as:

\begin{equation}
\frac{\Delta a}{a} = -\frac{\Delta M}{M},
\end{equation}

where $M$ is the total mass of the system, and $\Delta M$ is the mass lost.

When mass is lost from one of the inner binary components, both the inner and outer orbits of a triple expand. However, given the larger total mass of the outer binary (which includes the mass of the inner binary and the third companion), the relative expansion of the outer binary is smaller than that of the inner binary. Hence, the ratio of the outer to inner binary semi-major axes increases. Therefore, such mass loss drives a triple system to become less hierarchical.

When the mass loss is from the third companion, the opposite is true; only the outer binary expands, and the system becomes more hierarchical. The effect of mass exchange between the components is more complex, and the change in the hierarchy level of the system depends on the specific mass changes of the components.

While adiabatic mass loss does not directly affect eccentricities, the transition between different hierarchical regimes, where secular and dynamical evolution change eccentricities in different ways, results in 
an adiabatic mass loss significantly affecting the overall eccentricity evolution.

Mass loss and mass exchange therefore drive hierarchy changes in triple systems, which can be summarized with the following channels:
\begin{itemize}
    \item \textbf{Triple Evolution Dynamical Instability (TEDI):} This occurs when mass loss/transfer from the inner binary in evolving triples drives a hierarchical system to become unstable and follow chaotic evolution with its expected outcomes, such as triple disruptions or collisions/close interactions during the chaotic evolution.
    \item \textbf{Secular Freeze-out (SEFO):} In this scenario, mass is lost from the outer binary and/or transferred to the inner binary, causing an increase in the secular timescales, or even quenching of secular effects as the relative effects of GR precession become more important. In this case, the secular evolution slows down or even "freezes out".
    \item \textbf{Mass-loss/transfer-induced Eccentric vZLK (MIEK):} This occurs when mass loss/transfer changes the mass ratios in a triple system, such that the system transitions from evolving through regular vZLK evolution to eccentric vZLK.
    \item \textbf{Mass-loss-induced Quasi-Secular evolution (MIQS):} Similar to the TEDI case, mass loss/transfer from the inner binary drives the system to become less hierarchical, but in this case, it drives a system from the secularly evolving regime to the quasi-secular regime.
    \end{itemize}

Many of the important implications of stellar evolutionary effects have been summarized in several works \citep{Too+16,Too+20,Kum+23}, with some \citep{per+12,Ham+22,Too+22} focusing on the TEDI channel and resulting collisions and mergers through this channel.

As discussed earlier, secular and quasi-secular evolution can drive inner binary systems to strongly interact by exciting eccentricities and close pericenter interactions. Moreover, they allow for binary interactions that are inaccessible to isolated binary systems, as eccentricity evolution can produce different timings of strong interactions, effectively "jumping" over and delaying interactions to later evolutionary stages. Mass-loss and mass-transfer effects on secular and quasi-secular evolution further exacerbate these effects, allowing for different hierarchy and dynamical behavior to be triggered by stellar evolution and mass loss, and enabling post-main-sequence stars and compact objects to "jump" over evolutionary stages. For example, an inner binary star in a secularly evolving triple can be driven into high eccentricities and strongly interact. However, depending on the amplitude of eccentricity excitation driven by the triple configuration, it might not be sufficiently excited to interact even in this case. Mass loss can then drive the system to become less hierarchical and experience far larger eccentricity excitation, leading to the inner binary interacting when it otherwise would not have if it were not for the hierarchy change.

\textbf{Impact of prompt mass loss and natal kicks on orbital evolution:} Unlike the case of adiabatic mass loss, prompt mass loss, such as in the case of a supernova explosion, leads to an immediate change in both semi-major axis and eccentricity, and \emph{does} depend on the specific phases of the orbits at which the mass loss occurred.

When a massive star in a triple system undergoes a supernova explosion, the sudden loss of mass can impart a "Blaauw kick" \citep{bla61} due to the sudden change in the gravitational potential of the system. The potentially asymmetric nature of the explosion may also result in an additional "natal kick" to the resulting neutron star or black hole. These mass-loss and natal kicks can significantly alter the orbital configuration of the system \citep{Hil83}, often leading to the disruption of the triple system \citep{Koc21}. In some cases, the kick may eject one of the stars from the system entirely, while in other cases, the system may remain bound but with highly altered orbital parameters. Asymmetric kicks might also be more likely to change the mutual orbital inclinations of the orbits \citep{Pij+12}. Together, such natal kicks and/or prompt mass-loss processes can either unbind the triple or change its orbital configuration, thereby potentially changing its hierarchy and/or its type of dynamical evolution (secular/non-secular, unstable, etc.).

One should note that "prompt" mass loss might also occur from slower mass-loss processes, not only from supernova explosions. For example, common-envelope ejection introduces significant mass loss, which in this case affects the outer binary orbit and expands it, or possibly disrupts it \citep{Mic+19,Igo+20}, thereby allowing for the use of triples as probes to the mass-loss timescale of interacting binaries and other mass-loss processes. Even slow stellar evolutionary mass loss, which occurs on long, $\sim10^5$-year timescales, might be considered prompt for ultra-wide orbits with periods of this timescale.

We should note that the intermediate regime between the adiabatic and prompt mass-loss was less studied, but it introduces effects between these two types of interactions \citep[see also][and references therein]{Mic+19}.

\section{Fully coupled stellar evolution and dynamics}

As discussed above, triple stellar evolution entails the coupling of complex dynamics, arising for the first time only in such high-multiplicity systems, given that binary stars generally follow well-understood and simple non-chaotic Keplerian dynamics, and the evolution of single stars hardly involves kinematics (besides issues of natal kicks). Incorporating this non-trivial complex of physical processes together is a challenging endeavor. Exploring the vast phase space of initial conditions and possible evolutionary scenarios is difficult even for stellar binaries, while the much larger phase space and degrees of freedom for stellar triples make this even more challenging.

The development of population synthesis codes tailored for triple and higher-multiplicity stellar systems marks a significant leap in our ability to model complex stellar evolutionary scenarios. These advanced tools integrate the processes governing binary evolution with the distinct dynamical interactions present in triples and higher-order systems, offering a more holistic view of stellar population dynamics and their evolutionary trajectories. The first triple population synthesis code, which coupled dynamics and stellar evolution, was developed more than a decade ago. It incorporated stellar and inner binary evolution, and mass loss (albeit not including secular vZLK evolution), and pioneered the coupled stellar evolution and N-body dynamics for destabilized systems to explore mass-loss-induced instability in triples and binary-planet systems \citep{per+12}. More robust and sophisticated triple population synthesis codes have been developed since, based on similar ideas but far more advanced, with the first one in \citet{Ham+13} systematically incorporating coupled secular dynamics and dissipative processes. Currently, four primary population synthesis codes are utilized for modeling triple systems: MSE \citep{Ham+21}, TReS \citep{Too+16,Too+20}, TSE \citep{Ste+22}, and the code from \citeauthor{Naoz2016}'s group \citep[see][and references therein]{Sha+24}. While these codes share similar foundational approaches, they differ in several aspects and capabilities.

These codes incorporate stellar evolution prescriptions for single and binary stars and couple them to the dynamical evolution of triples. They account for simplified mass loss and mass transfer, inner binary common envelope, orbital evolution, and the chaotic dynamics inherent in three-body systems, integrating these with secular evolution processes modeled through secular Hamiltonian dynamics for secular evolution and transitioning to N-body codes for chaotic and quasi-secular evolution.

The codes tackle the challenge of managing processes that operate on vastly different timescales through adaptive time-stepping and the selective application of integration methods as required. However, they diverge in key areas, including the transition criteria between dynamical regimes, the stellar evolution prescriptions employed (SSE/BSE vs. SeBa), the treatment of eccentric mass transfer and triple mass transfer, and the types of N-body codes used. They also differ in the methods for coupling dynamical and stellar evolutionary processes.

As in the case of binary evolution, triple population synthesis models are constructed to probe the overall impact of the various evolutionary channels on the outcomes of realistic ensembles of stellar systems, to compare them with observed populations, and thereby constrain the physical processes and assumptions of the triple stellar evolutionary models. They can also potentially predict and explain the observed properties of stellar systems and astrophysical phenomena arising from their evolution, including the various explosive transients they produce, the production of elements in the universe, and the implications for galaxy evolution and feedback processes.

Making use of population synthesis studies and overall understanding, main triple evolutionary channels become apparent \citep{Mic+14,Too+16,Too+20,Kum+23}, and the picture that emerges is that triples play a key role in the evolution of stars and that a major part of stellar astrophysical phenomena is many times dominated by the effects of triple stellar evolution. This suggests that much of the previous studies of stellar evolution and stellar populations, which did not account for triples, potentially provide a very partial and sometimes incorrect view and understanding of stellar populations, their constituents, and their evolution. This is particularly true for massive stars, the majority of which form in triple (or higher multiplicity) systems. Thereby triple evolution likely plays a major role in the production and evolution of explosive transients and X-ray binaries and in determining their rates/frequencies. These include both core-collapse and thermonuclear supernovae \citep{Ham+13,Too+18,Raj+23,Sha+23}, neutron star and black hole mergers and binary properties, and thereby also GW sources and gamma-ray bursts \citep{Ham+19b,Ham+19c,Ant+17,Sil+17,Rod+18,Ste+22,Ste+22b,Vig+21,Mar+22,Gen+24}, as well as micro-tidal disruption events \citep{per+16}, stellar mergers \citep{per+09b,per+12,Nao+14,Ham+22b,Too+22,Sha+24}, and interacting binaries \citep{Ham+22b,Sha+24b}. It also affects the production of various types of stars that cannot arise from regular single stellar evolution \citep{Pre+22,Gao+23}.

\section{Stellar triple evolution and dynamics}
The evolution of single isolated stars has been explored for more than a century and is the most fundamental aspect of astrophysics and one of the most well-understood fields in astrophysics (while still holding many open questions). All aspects of single stellar evolution (and its open issues) are directly transferred to and affect the evolution of binary stars. Binary stellar evolution has been extensively studied for more than 50 years and has introduced many new physical processes and phenomena.

A large fraction of all stars form and/or reside in binary systems, many of which are interacting systems. Therefore, many current astrophysical phenomena, and in particular explosive phenomena arising from various types of binary mergers/mass-transfer, and a wide variety of stellar phenomena cannot be produced through single stellar evolution. Hence, binary stellar evolution has become another pillar of astrophysics. Compared with single stellar evolution, many aspects dramatically affecting binary stellar evolution are still not understood, including highly debated fundamental issues such as mass transfer and common-envelope evolution, as well as natal kicks of compact objects (an issue not well understood even in single stellar evolution), which have wide implications for the evolution of compact-object binaries and interacting binaries.

For many decades, triple systems received far less attention than single or binary stellar systems. This is likely in part due to the difficulty in detecting, identifying, and characterizing such systems, and partly due to the difficulty in modeling such systems. The former led to slow progress in assessing their frequency. Only over the last two decades has it become progressively clear that triple systems are highly abundant (see the observations section) and that the majority of massive stars form in triple (or even higher multiplicity) systems. Like binary evolution inheriting the understanding and open issues of single stellar evolution, triple stellar evolution makes use of all the understanding regarding binary stellar evolution but also inherits its open questions and issues. Similar to how binary evolution introduced completely new physical processes and evolutionary channels, triple evolution introduced complex dynamics and its coupling to stellar evolution, giving rise to new processes and evolutionary channels. In particular, triple systems inherently involve the dynamical issues of the three-body problem.

The three-body problem and the dynamics of three gravitating bodies is one of the oldest studied problems in astrophysics, beginning with Newton and the study of the Sun-Earth-Moon system, through the foundation of chaos theory arising from the study of the three-body problem, to modern times and the characterization and understanding of secular processes, as well as modern tools and studies of the chaotic three-body problem, from numerical simulations to statistical physics approaches. As discussed above, triple stellar evolution involves the coupling of the complex dynamics of triples with the complex evolution of stars and binaries. Naturally, the understanding of triple stellar evolution is therefore far more difficult, and it is still in its early stages. However, it is already becoming clear, given the frequency of triple systems and the fundamental differences between binary and triple stellar evolution, that triple stellar evolution is another pillar of stellar evolution and astrophysics at large, and serves as the gateway to understanding even higher multiplicity systems.

\section{Summary and Outlook}

Triple stellar systems represent a fundamental component of stellar populations that requires consideration of both complex dynamical processes and stellar evolution. This chapter has presented the current understanding of these systems, from their observational properties to the physical processes that govern their evolution. The key aspects covered include:
\begin{itemize}
    \item The abundance of triple systems, particularly among massive stars, where they constitute the majority of stellar systems, necessitates their inclusion in our understanding of stellar evolution and stellar populations. 
    \item The evolution of triples differs fundamentally from that of binary systems due to the introduction of complex dynamical processes unique to three-body systems, including secular von Zeipel-Lidov-Kozai evolution, eccentric vZLK processes, and quasi-secular evolution, as well as potentially chaotic dynamics.
    \item The coupling between these dynamical processes and stellar evolution, including mass loss, mass transfer, and dissipative processes, leads to unique evolutionary pathways inaccessible to binary systems. This coupling introduces new channels for the formation of compact and interacting binaries, merger products, and various types of explosive phenomena.
    \item Future advancements in observational capabilities, particularly with upcoming facilities, will improve our statistical understanding of triple systems. Continued development of theoretical frameworks and computational capabilities will enable more detailed modeling of their evolution. These developments are essential for advancing our understanding of stellar populations and their outcomes, particularly for massive stars where triple evolution likely dominates their evolutionary pathways.
  \end{itemize}
  
  The study of triple stellar systems thus represents a crucial component in our understanding of stellar evolution, with broad implications for various astrophysical phenomena, from the formation of compact object binaries to the production of explosive transients and gravitational wave sources.

\begin{ack}[Acknowledgments]
HBP acknowledges support from ERC consolidator grant "SNEX" and would like to thank G. H. Bashkar for his assistance.
\end{ack}
  
\bibliographystyle{agsm}

\end{document}